\documentclass[sigconf,10pt]{acmart} 

\usepackage[T1]{fontenc}
\usepackage{booktabs}
\usepackage{multirow, array}
\usepackage{amsmath}
\usepackage{siunitx}
\usepackage{subcaption}
\usepackage{csvsimple}
\usepackage{graphicx}
\graphicspath{{assets/figures/}{assets/logos/}}
\usepackage{booktabs, multirow, xcolor, makecell}
\usepackage{tabularx}
\newcolumntype{Y}{>{\raggedright\arraybackslash}X}
\usepackage[normalem]{ulem}

\newcommand{\cls}{\ensuremath{^{\mathrm{C}}}}
\newcommand{\reg}{\ensuremath{^{\mathrm{R}}}}

\newif\ifannotated
\newcommand{\new}[1]{\ifannotated \textcolor{blue}{#1}\relax \else #1\relax \fi}
\newcommand{\remove}[1]{\ifannotated {\color{red}\sout{#1}}\relax \fi}

\annotatedfalse
\title{Breathing and Semantic Pause Detection and Exertion-Level Monitoring in Post-Exercise Speech}

\author{Yuyu Wang, Wuyue Xia, Huaxiu Yao, and Jingping Nie}
\affiliation{\institution{University of North Carolina at Chapel Hill, Chapel Hill, NC} \country{\{yuyuwang, wuyuexia, jingping\}@unc.edu, huaxiu@cs.unc.edu}}
\renewcommand{\shortauthors}{Wang et al.}

\acmYear{2025}\copyrightyear{2025}
\acmConference[IASA '25]{3rd ACM International Workshop on Intelligent Acoustic Systems and Applications}{November 4--8, 2025}{Hong Kong, China}
\acmBooktitle{3rd ACM International Workshop on Intelligent Acoustic Systems and Applications (IASA '25), November 4--8, 2025, Hong Kong, China}
\acmDOI{10.1145/3737901.3768369}
\acmISBN{979-8-4007-1978-3/25/11}

\acmYear{2025}\copyrightyear{2025}
\acmConference[IASA '25]{3rd ACM International Workshop on Intelligent Acoustic Systems and Applications}{November 4--8, 2025}{Hong Kong, China}
\acmBooktitle{3rd ACM International Workshop on Intelligent Acoustic Systems and Applications (IASA '25), November 4--8, 2025, Hong Kong, China}
\acmDOI{10.1145/3737901.3768369}
\acmISBN{979-8-4007-1978-3/25/11}

\copyrightyear{2025}


\copyrightyear{2025}



\begin{document}
\title{Breathing and Semantic Pause Detection and Exertion-Level Classification in Post-Exercise Speech}

\begin{abstract}
Post-exercise speech contains rich physiological and linguistic cues, often marked by semantic pauses, breathing pauses, and combined breathing–semantic pauses. Detecting these events enables assessment of recovery rate, lung function, and exertion-related abnormalities. However, existing works on identifying and distinguishing different types of pauses in this context are limited. In this work, building on a recently released dataset with synchronized audio and respiration signals, we provide systematic annotations of pause types. Using these annotations, we systematically conduct exploratory \emph{breathing and semantic pause detection} and \emph{exertion-level classification} across deep learning models (GRU, 1D CNN-LSTM, AlexNet, VGG16), acoustic features (MFCC, MFB), and layer-stratified Wav2Vec2 representations. We evaluate three setups—single feature, feature fusion, and a two-stage detection–classification cascade—under both classification and regression formulations. Results show per-type detection accuracy up to 89\% for semantic, 55\% for breathing, 86\% for combined pauses, and 73\% overall, while exertion-level classification achieves 90.5\% accuracy, outperforming prior work. 

\end{abstract}

\maketitle
\vspace{-0.1in}
\section{Introduction}\label{sec:introduction}

Post-exercise speech carries both physiological and linguistic cues, often marked by distinct pauses, micro-breaths, or even exercise-induced wheezing or asthma~\cite{nie2025multi, mitra2024pre}. These pauses can be categorized as semantic pauses, which occur at linguistic boundaries, breathing pauses, which reflect increased respiratory demand after exercise or due to dyspnoea~\cite{smoliga2016common}, or combined breathing-semantic pauses, where both co-occur. Tracking these patterns helps assess recovery rate, lung function, and potential respiratory abnormalities~\cite{ossewaarde2025role}.

A wide range of acoustic features and models have been employed in speech and bioacoustic analysis. Traditional features such as Mel filter banks (MFBs), Mel-frequency cepstral coefficients (MFCCs), and power spectral density (PSD) remain widely used in speech and body-sound tasks~\cite{wang2024turbocharge, nie2025multi}. More recently, representations from pre-trained self-supervised speech foundation models (FMs) have shown superior performance in tasks such as emotion recognition and cardiorespiratory sound analysis, compared to handcrafted acoustic features~\cite{mitra2024investigating, nie2025foundation}. These self-supervised speech FMs, such as Wav2Vec 2.0 (W2V2) and HuBERT, provide layer-stratified representations; mid layers tend to carry paralinguistic information while upper layers skew toward linguistic semantics~\cite{baevski2020wav2vec, hsu2021hubert, pasad2021layer}. 

In terms of model architectures, widely adopted deep learning models have also shown strong performance across speech and physiological signal analysis tasks. \citeauthor{mitra2024pre} applied a Conv-LSTM with W2V2 representations to uncover breathing patterns in speech and estimate respiratory rate (RR) \cite{mitra2024pre}. Modified 2D CNNs have been effective for heart rate and heart murmur detection from phonocardiograms~\cite{nie2024model}, while VGG16 has demonstrated strong performance in respiratory sound classification, including the detection of crackles, wheezes, and rhonchi~\cite{kim2021respiratory}.

However, most existing speech or cardiorespiratory sound analysis systems are developed using resting-state speech or controlled corpora\new{, and do not incorporate semantic information, instead, they use methods such as adaptive complementary decomposition with IMF-energy thresholds~\cite{alimuradov2017speech}}. In contrast, post-exercise speech introduces irregular breathing rhythms, micro-breaths, overlapping speech–breathing events, motion artifacts, and device-handling noise. A recent dataset capturing speech, breathing, and phonocardiograms under exertion levels provides standardized data and protocols~\cite{nie2025multi} and includes a preliminary analysis to showing that self-reported exertion level can be decoded from the post-exercise speech. Moreover, breathing-only pauses occur primarily at high exertion, where they are intrinsically harder to detect due to being short, low in signal-to-noise ratio, and relatively sparse, resulting in both intrinsic detection difficulties and data imbalance in the dataset~\cite{nie2025multi}. These challenges underscore the need for systematic annotations and benchmarking of breathing and semantic pause detection and exertion-level classification in post-exercise speech.

Considering the aforementioned opportunities and limitations, we manually annotate the onsets of semantic pauses (\textbf{\texttt{S}}), breathing pauses (\textbf{\texttt{B}}), and combined breathing-semantic pauses (\textbf{\texttt{BS}}), labeling all remaining segments as (\textbf{\texttt{O}}), for the post-exercise reading and spontaneous speech data frome \cite{nie2025multi}, using both audio and chest-belt respiration signals as references (see Section~\ref{secsec:data_annotation}). These annotations, which will be open-sourced to facilitate research in related areas, enable systematic benchmarking of both \textbf{breathing and semantic pause detection} and \textbf{exertion-level classification}. To this end, we conduct extensive exploratory studies with (\textit{i}) widely adopted deep learning (DL) models for body-sound analysis (GRU, 1D CNN-LSTM, AlexNet, VGG16) and (\textit{ii}) acoustic features (MFCC, MFB) combined with 4th-, 6th-, and 12th-layer representations from the pre-trained W2V2-base encoder. Our evaluation covers three setups: \textcircled{1} DL models with a single feature, \textcircled{2} DL models with feature fusion, and \textcircled{3} a two-stage approach that first detects pause activity and then classifies pause type. We benchmark breathing and semantic pause detection under all three setups in two tasks, classification and regression, and evaluate exertion-level classification under setups \textcircled{1} and \textcircled{2}, providing the first comprehensive benchmark for post-exercise speech analysis. Across the three DL model setups, per-class accuracy reaches up to 89\% for \textbf{\texttt{S}}, 55\% for \textbf{\texttt{B}}, 86\% for \textbf{\texttt{BS}}, and 73\% for overall accuracy. The exertion level prediction results in 90.48$\%$ of accuracy, outperforming exsiting work~\cite{nie2025multi}.

\section{Method}\label{secsec:data_annotation}
\begin{figure}[t]
    \centering
    \caption{An example waveform of a 15-s audio snippet with annotated pause regions.}
    \label{fig:example_label}
    \includegraphics[width=\linewidth]{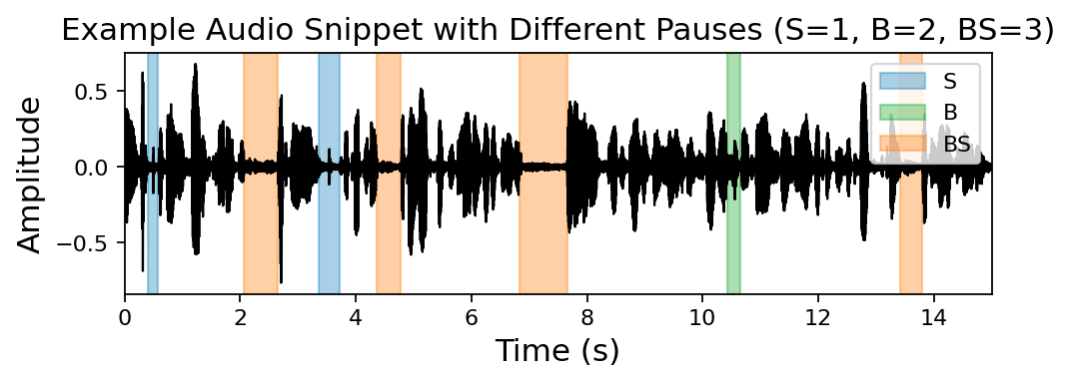}
\end{figure}

\enlargethispage{\baselineskip}
The dataset ~\cite{nie2025multi} includes multiple modalities (audio, respiration, phonocardiograms, etc.). This study performed frame-wise annotation on two subsets: (\textit{i}) \textbf{reading} (participants read from a provided paragraph list) and (\textit{ii}) \textbf{spontaneous speech} (participants spoke freely). As shown in Figure~\ref{fig:example_label}, each recording was annotated frame-wise with one of four labels: \textbf{\texttt{S}}, \textbf{\texttt{B}}, \textbf{\texttt{BS}}, or \textbf{\texttt{O}}. \remove{Annotations were independently performed by three annotators, and final labels were determined by majority vote.}\new{Three annotators labeled independently; final labels were determined by majority vote.}

\begin{figure*}[t]
    \centering
    \begin{subfigure}[t]{0.32\textwidth}
        \centering
        \includegraphics[width=\linewidth]{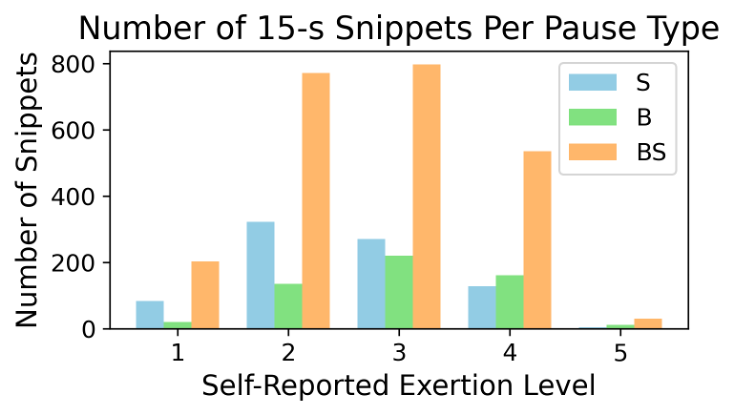}
    \end{subfigure}
    \hfill
    \begin{subfigure}[t]{0.32\textwidth}
        \centering
        \includegraphics[width=\linewidth]{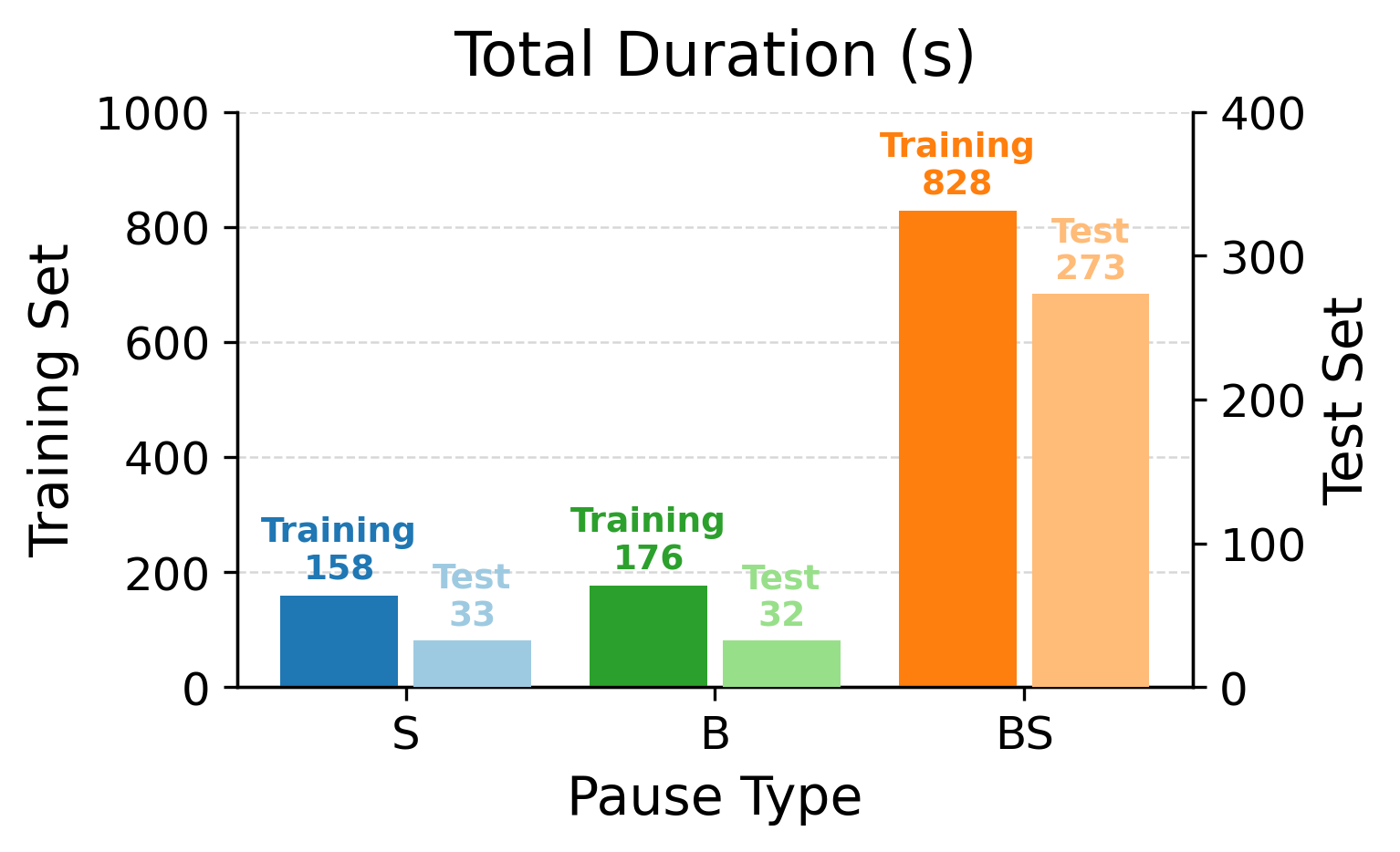}
    \end{subfigure}
    \hfill
    \begin{subfigure}[t]{0.32\textwidth}
        \centering
        \includegraphics[width=\linewidth]{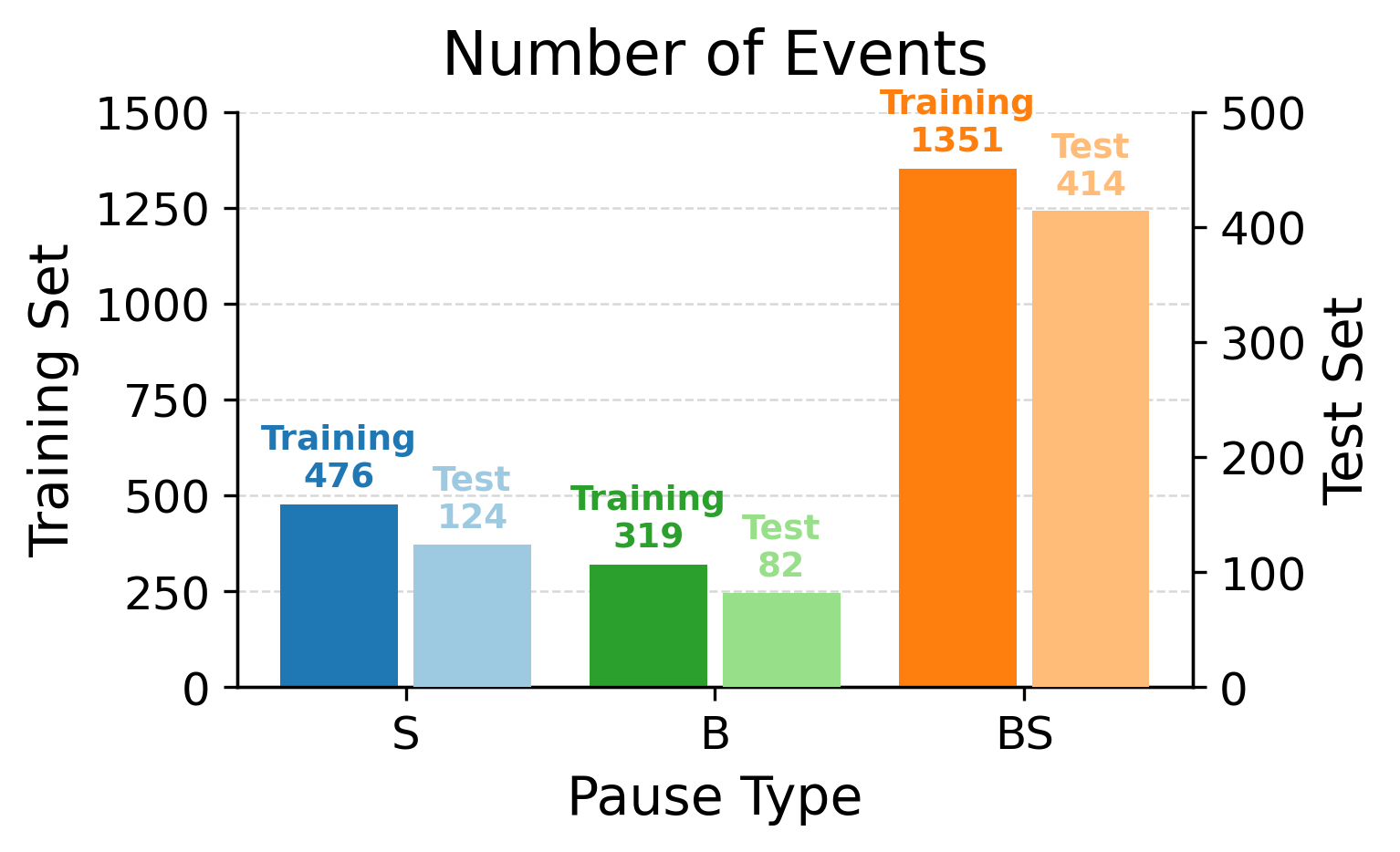}
        \end{subfigure}
    \caption{Comparison of exertion counts, duration distribution, and event numbers across train and test sets.}
    \label{fig:dataset_stats}
    \vspace{-0.15in}
\end{figure*}

\subsection{Features}
\noindent\textbf{Acoustic Features:} 
Post-exercise speech recordings in the dataset~\cite{nie2025multi} were downsampled to $16\thinspace kHz$ and mean–variance normalized. MFBs and MFCCs were extracted at 50 Hz from normalized 15 s audio snippets, aligned with frame-wise labels to form 750-frame sequences of 40 bands or coefficients.

\noindent\textbf{FM Embedding:}
The W2V2-base model ~\cite{baevski2020wav2vec} was employed as an additional feature, which was pre-trained self-supervised on 960 hours of Librispeech audio and consists of 12 transformer layers with 768-dimensional hidden representations. The model parameters were kept frozen, and representations were extracted from the 4th, 6th, and 12th audio encoder layers (Emb-4, Emb-6, and Emb-12). The embeddings are resampled to $50\thinspace Hz$. This sampling rate strikes a balance between keeping most of W2V2's information and a tidy downsampling from the original audio, enabling the frame-wise concatenation of acoustic features and embeddings. 

\subsection{Model Training}  
\noindent \textbf{Pause Detection:} \{\textbf{\texttt{O}, \texttt{S}, \texttt{B}, \texttt{BS}}\} are denoted as \{0, 1, 2, 3\} in both classification (C) and regression (R) tasks. Let $N$ be the number of samples (segments) and $T$ the number of frames per sample. Predictions and targets are denoted by $\hat y_{it}$ and $y_{it}$. The cross-entropy (CE) loss was used for classification, whereas for regression tasks, we employed the Huber loss:  
\begin{equation}
\setlength{\abovedisplayskip}{6pt}
\setlength{\abovedisplayshortskip}{0pt}
\setlength{\belowdisplayskip}{6pt}
\setlength{\belowdisplayshortskip}{6pt}
\mathcal{L}_{\mathrm{Huber}} = \frac{1}{NT}\sum_{i,t} 
\begin{cases}
\tfrac{1}{2} e_{it}^{2}, & |e_{it}|\le\delta,\\[2pt]
\delta\bigl(|e_{it}|-\tfrac{\delta}{2}\bigr), & |e_{it}|>\delta,
\end{cases}
\quad e_{it}=\hat y_{it}-y_{it},
\label{eq:huber_loss}
\end{equation}
where $e_{it}$ is the prediction error.

The two-stage setup used binary cross-entropy (BCE) loss in Stage~1 for pause detection and a duration-aware focal (DAF) loss in Stage~2:  
\begin{equation}
\setlength{\abovedisplayskip}{4pt}
\setlength{\abovedisplayshortskip}{0pt}
\setlength{\belowdisplayskip}{4pt}
\setlength{\belowdisplayshortskip}{6pt}
\mathcal{L}_{\mathrm{DAF}} = \frac{1}{NT}\sum_{i,t}\alpha\,w_{c_{it}}\left(\frac{|e_{it}|}{\delta}\right)^{\gamma}\mathrm{Huber}_{\delta}(e_{it}),
\label{eq:daf_loss}
\end{equation}
where $w_{c_{it}}$ denotes class-dependent weights, $\alpha$ is a global scaling factor that balances the regression term with other objectives, $\gamma$ controls the focal strength, increasing the relative emphasis on harder examples (\textbf{\texttt{S}} and \textbf{\texttt{B}})  with larger errors, and $\delta$ determines the sensitivity to outliers.

\noindent \textbf{Exertion Level Classification:} Similar to \cite{nie2025multi}, the five-class exertion- level labels (levels 1–5) were clustered to binary classes (3-5 as High vs. 1-2 as Low). COnsistent RAnk Logits (CORAL) output layer was adopted as the output layer instead of a traditional classifier~\cite{cao2019consistent}. 

\begingroup
\setlength{\abovedisplayskip}{6pt}
\setlength{\abovedisplayshortskip}{0pt}
\setlength{\belowdisplayskip}{6pt}
\setlength{\belowdisplayshortskip}{6pt}
\begin{equation}
    \mathcal{L}_{\mathrm{CORAL}}=-\frac{1}{N}\sum_{i=1}^N \sum_{k=1}^{K-1}[t_{ik}\log(p_{ik})+(1-t_{ik})\log (1-p_{ik})]
    \label{eq:coral}
\end{equation}
\endgroup
where $K$ is the number of classes, $t_{ik}$ is the ordinal ground truth, and $p_{ik}$ is the predicted probability. All models were trained with a mini-batch size of 64, using Adam optimizer with an initial learning rate of 0.0001.  

\subsection{Data Preparation} Recordings with abnormal characteristics (e.g., mismatched or missing respiration data) were excluded, leaving 307 audio files; recordings shorter than 15 s were further discarded, resulting in 296 valid audio files. These were split into training (70\%), validation (15\%), and test (15\%) sets with balanced duration, \new{with no subject overlap across splits}. Mean variance normalization was applied to the time-domain audio signal to mitigate the variability across subjects and data collection environments. Training and validation files were segmented using a 15-s sliding window with 1 s stride, yielding frame-wise sequences of 750 time steps (50 Hz). \new{2404, 222, and 94 15-s audio snippets in the training, validation, and test sets were generated from the 296 audio files, respectively.} Figure~\ref{fig:dataset_stats} shows the distribution of duration and pause events across the training and test sets.

\subsection{Three Setups for Pause Detection}\label{secsec: DL Model with Single Feature}
The set of four DL models explored in the three setups includes GRU, 1D CNN-LSTM, AlexNet, and VGG16; each model is applied to both classification and regression formulations of pause type prediction \new{with only the output layer altered, ensuring a fair comparison}. The GRU model consists of two bidirectional layers, while the 1D CNN-LSTM applies a temporal convolutional layer followed by a two-layer bidirectional LSTM. The AlexNet variant contains five convolutional layers, and the VGG16 model is composed of four multi-conv-pooling blocks. Temporal pooling and fully connected layers are applied at the end of AlexNet and VGG16 for frame-wise classification/regression. 

\begin{figure}[t!]
    \centering
    \includegraphics[width=0.95\columnwidth]{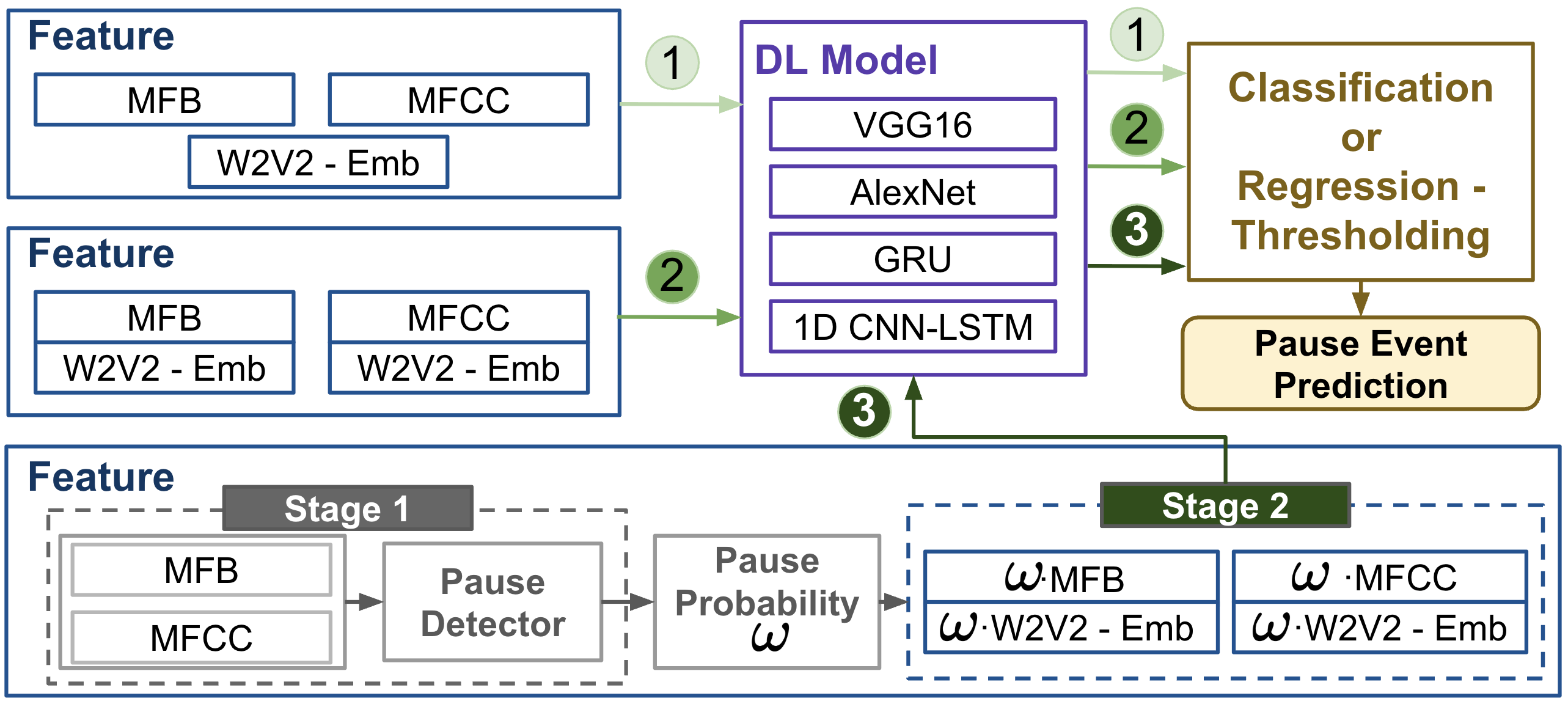}
    \caption{The workflow for Setup \textcircled{1}, \textcircled{2}, and \textcircled{3}.}
    \label{fig:setup}
\end{figure}

In the single feature setup, the input is 40- or 768-dim, depending on the feature, while in the fused setup it is concatenated to 808-dim. All inputs are in a sequence of length 750.
Let $T{=}750$ denote the number of frames. Depending on the setup, the model consumes either a single feature $X\in\mathbb{R}^{T\times F}$ (Setup~\textcircled{1}; MFB/MFCC: $F{=}40$, W2V2 Emb-4/Emb-6/Emb-12: $F{=}768$) or a fused feature $X=[A;E]\in\mathbb{R}^{T\times(F_A+F_E)}$ (Setups~\textcircled{2},\textcircled{3}), where $A\in\mathbb{R}^{T\times 40}$ is an acoustic feature, MFB or MFCC, and $E\in\mathbb{R}^{T\times 768}$ is a W2V2 embedding, outputting the final result, $Y \in \mathbb{R}^{T \times 1}$, a frame-wise label for classification or a scalar pause type score in [0,3] for regression. Both classification and regression frame-wise predictions are post-processed (see Section \ref{secsec:post_processing}) to form valid pause event predictions for model performance evaluation. The overall workflow of the three setups is illustrated in Figure~\ref{fig:setup}.

\textbf{\textcircled{1} DL Model with Single Feature:} As indicated by \textcircled{1} in Figure~\ref{fig:setup}, a single feature, MFB, MFCC, or W2V2 embedding, is provided as input to each of the four DL models, which outputs frame-wise predictions of pause types. 

\textbf{\textcircled{2} DL Model with Feature Fusion:} An acoustic feature $A$ is fused with a W2V2 embedding $E$ by vertical concatenation, and the fused representation $X=[A;E]\in\mathbb{R}^{T\times808}$ is then passed into a DL model to produce frame-wise predictions, as shown in Figure~\ref{fig:setup}. 

\textbf{\textcircled{3} Two-Stage Approach:}
As denoted in Figure~\ref{fig:setup}, a two-stage pipeline is used in this setup. In Stage~1, a pause detector estimates frame-wise pause probabilities $\boldsymbol{\omega}\in[0,1]^T$ from an acoustic feature $A$ and uses them to re-weight both $A$ (the same $A$ used to obtain $\boldsymbol{\omega}$) and a W2V2 embedding $E$, i.e., $\tilde A=\boldsymbol{\omega}\odot A$ and $\tilde E=\boldsymbol{\omega}\odot E$. Stage~2 concatenates the weighted representation, $X=[\tilde A;\tilde E]\in\mathbb{R}^{T\times 808}$, and feeds $X$ to a DL model to produce frame-wise outputs. Stage~1 employs a two-layer bidirectional LSTM (128 units per direction) as the pause detector since it achieved the highest pause/VAD accuracy (0.94) on our data, outperforming a simple MLP (0.89) and an RMS-energy baseline (0.24). 

    \begin{figure}[t!]
        \centering
        \caption{Left: Regression prediction sequences averaged and transformed in frequency space, in order to decide on the best low-pass cutoff frequency. Right: Comparison of ground-truth pauses with regression predictions across different post-processing stages.}
        \includegraphics[width=\linewidth]{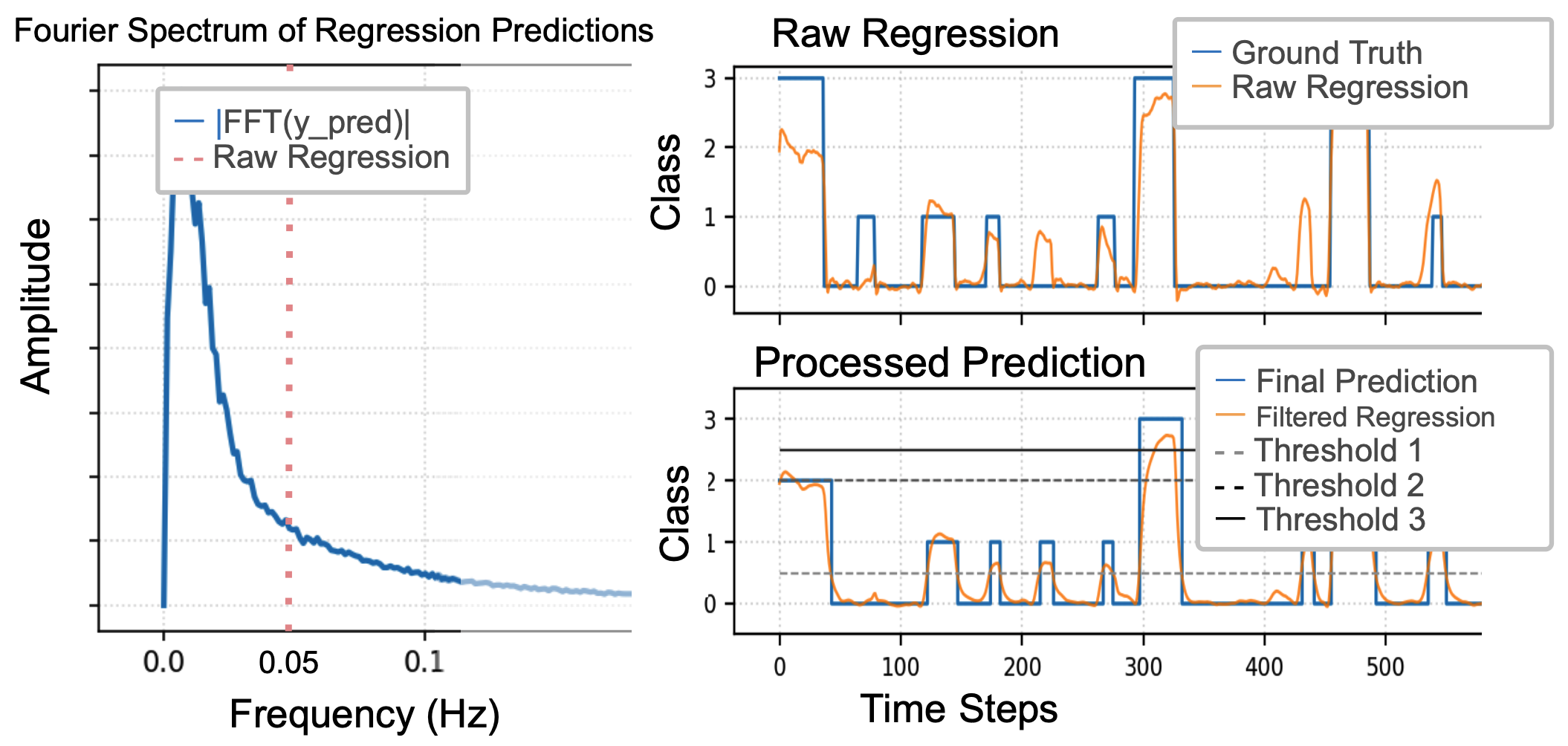}
        \label{fig:postprocess}
        \vspace{-0.1 in}
    \end{figure}
    
\subsubsection{Post Processing:}\label{secsec:post_processing}
To mitigate the noise in the \remove{predictions in the }\underline{\textit{regression}} \new{predictions}\remove{task}, we applied a low-pass filter with a cutoff frequency of $0.05\thinspace Hz$. As shown in Figure~\ref{fig:postprocess}, the averaged Fourier transform spectra for the regression predictions reveal an "elbow" at $0.05\thinspace Hz$, beyond which the amplitude decays and higher frequencies primarily represent noise. After filtering, adjacent low-level predictions (1–2) were merged with neighboring higher-level predictions (2–3), reflecting the continuity of natural pauses. We swept thresholds across model outputs to optimally map regression predictions to classes. Figure~\ref{fig:postprocess} shows a representative result, with outputs suitable for pause categorization and comparable to classification setups. The \remove{predictions in the }\underline{\textit{classification}} \new{predictions}\remove{task} were cleaned by enforcing a minimum event length of 3 frames, bridging brief zero gaps between identical labels, and unifying each contiguous non-zero segment to its majority label.

\section{Results}
\begin{table}[t]
  \centering
  \setlength{\tabcolsep}{4pt}
  \small
  \caption{Representative per-type accuracies (\textbf{\texttt{S}}, \textbf{\texttt{B}}, \textbf{\texttt{BS}}) and overall results across three DL model setups, where \cls\ and \reg\ indicate \emph{classification} and \emph{regression} tasks.}
  \label{tab:pause detection results}
  \begin{tabularx}{\linewidth}{l X c c c c}
    \toprule
    \textbf{Model} & \textbf{Feature(s)} & \textbf{\texttt{S}} & \textbf{\texttt{B}} & \textbf{\texttt{BS}} & \textbf{Overall} \\
    \midrule

    \multicolumn{6}{l}{\textbf{\textcircled{1} Model - Single Feature}} \\ 
    1D CNN-LSTM\reg & Emb-6   & 0.65 & 0.23 & \textcolor{red}{0.86} & \textcolor{red}{0.73} \\
    1D CNN-LSTM\reg & MFB     & 0.67 & 0.27 & 0.81 & 0.71 \\
    GRU\reg         & Emb-4   & 0.44 & \textcolor{red}{0.55} & 0.68 & 0.61 \\
    GRU\cls         & MFB     & \textcolor{red}{0.80} & 0.39 & 0.57 & 0.59 \\
    VGG16\cls       & Emb-4   & \textcolor{red}{0.80} & 0.07 & 0.68 & 0.62 \\
    VGG16\cls       & MFCC    & 0.49 & 0.24 & 0.74 & 0.62 \\
    AlexNet\reg     & Emb-12  & 0.03 & 0.01 & 0.23 & 0.16 \\
    AlexNet\reg     & MFB     & 0.63 & 0.06 & 0.67 & 0.57 \\
    \midrule

    \multicolumn{6}{l}{\textbf{\textcircled{2} Model - Feature Fusion}} \\ 
    1D CNN-LSTM\reg & MFB+Emb-4   & 0.69 & \textcolor{red}{0.48} & 0.69 & 0.66 \\
    1D CNN-LSTM\reg & MFB+Emb-6   & 0.76 & 0.16 & 0.81 & \textcolor{red}{0.71} \\
    1D CNN-LSTM\reg & MFCC+Emb-4  & \textcolor{red}{0.80} & 0.35 & 0.75 & 0.70 \\
    1D CNN-LSTM\reg & MFCC+Emb-6  & 0.69 & 0.39 & 0.73 & 0.67 \\
    GRU\reg         & MFB+Emb-4   & 0.44 & 0.45 & 0.76 & 0.65 \\
    GRU\cls         & MFB+Emb-6   & 0.63 & \textcolor{red}{0.48} & 0.69 & 0.65 \\
    GRU\cls         & MFCC+Emb-4  & 0.58 & 0.27 & \textcolor{red}{0.82} & 0.69 \\
    GRU\reg         & MFCC+Emb-6  & 0.54 & 0.33 & 0.65 & 0.58 \\
    \midrule

    \multicolumn{6}{l}{\textbf{\textcircled{3} Two-Stage Setup}} \\ 
    1D CNN-LSTM\reg & MFB+Emb-4   & 0.62 & 0.34 & 0.84 & \textcolor{red}{0.72} \\
    1D CNN-LSTM\cls & MFB+Emb-6   & 0.79 & 0.51 & 0.62 & 0.64 \\
    1D CNN-LSTM\reg & MFCC+Emb-4  & 0.59 & 0.33 & 0.79 & 0.68 \\
    1D CNN-LSTM\cls & MFCC+Emb-6  & 0.72 & \textcolor{red}{0.54} & 0.58 & 0.61 \\
    GRU\cls         & MFB+Emb-4   & \textcolor{red}{0.89} & 0.43 & 0.49 & 0.57 \\
    GRU\cls         & MFB+Emb-6   & 0.85 & 0.41 & 0.61 & 0.63 \\
    GRU\reg         & MFCC+Emb-4  & 0.51 & 0.13 & \textcolor{red}{0.85} & 0.68 \\
    GRU\cls         & MFCC+Emb-6  & 0.68 & 0.48 & 0.48 & 0.52 \\
    \bottomrule
  \end{tabularx}
\end{table}

\begin{table*}[t!]
    \small
    \centering
    \caption{Exertion level prediction accuracy of models and input configurations on new data format. Comparison baselines are marked in blue, while those that achieved significant improvements are marked in red.}
    \label{tab:benchmark}
    \setlength{\tabcolsep}{5.5pt}
    \begin{tabular}{ll*{12}{c}}
        \toprule
        \textbf{Subset} & \textbf{Layer}
            & \multicolumn{3}{c}{VGG16} & \multicolumn{3}{c}{AlexNet} & \multicolumn{3}{c}{GRU} & \multicolumn{3}{c}{1D CNN-LSTM} \\
        \cmidrule(lr){3-5}\cmidrule(lr){6-8}\cmidrule(lr){9-11}\cmidrule(lr){12-14}
            & 
            & MFB & MFCC & W2V2 & MFB & MFCC & W2V2 & MFB & MFCC & W2V2 & MFB & MFCC & W2V2 \\
        \midrule
        \multirow{4}{5em}{\textbf{Spontaneous Speech}}
            & \textbf{No Emb}
                & 0.5238 & \textcolor{red}{0.8571} & N.A. & 0.6667 & 0.8095 & N.A. & 0.6667 & \textcolor{red}{0.9048} & N.A. & 0.6667 & 0.8095 &  N.A.\\
            & \textbf{Emb-4}
                & 0.5238 & 0.7619 & 0.3810 & 0.7619 & 0.5238 & 0.8095 & 0.7143 & 0.6667 & 0.7143 & 0.8571 & 0.8095 & \textcolor{blue}{0.7619} \\
            & \textbf{Emb-6}
                & 0.5238 & 0.7619 & 0.3810 & 0.5238 & 0.7143 & 0.7143 & \textcolor{red}{0.9048} & 0.7143 & 0.8571 & 0.8095 & 0.8095 & 0.8571 \\
            & \textbf{Emb-12}
                & 0.6667 & 0.6667 & 0.3810 & 0.5714 & \textcolor{red}{0.9048} & 0.4762 & 0.7619 & 0.6667 & 0.8095 & 0.8571 & 0.7143 & 0.7619 \\
        \midrule
        \multirow{4}{5em}{\textbf{Spontaneous Speech \& Reading}}
            & \textbf{No Emb}
                & 0.7340 & 0.8298 & N.A. & 0.7340 & 0.7660 & N.A. & 0.7447 & 0.7553 & N.A. & 0.6383 & 0.7979 & N.A. \\
            & \textbf{Emb-4}
                & \textcolor{red}{0.8617} & 0.7021 & 0.5000 & 0.6383 & 0.6702 & 0.5745 & 0.7979 & 0.6915 & 0.7128 & 0.7660 & 0.7766 & \textcolor{blue}{0.8191} \\
            & \textbf{Emb-6}
                & 0.7340 & 0.7447 & 0.5213 & 0.6064 & 0.5000 & 0.6702 & 0.7660 & 0.7340 & 0.7021 & 0.8298 & 0.7872 & 0.8511 \\
            & \textbf{Emb-12}
                & 0.7128 & 0.7128 & 0.4894 & 0.6489 & 0.7660 & 0.7447 & 0.7872 & 0.7340 & 0.5957 & 0.7553 & 0.7660 & 0.7872\\
        \bottomrule
    \end{tabular}
    \vspace{-0.1in}
\end{table*}

\noindent\textbf{Breathing and Semantic Pause Detection:} 
Evaluation metrics for both regression and classification predictions include per-type accuracy and overall event detection accuracy. Events were extracted as contiguous non-zero segments from the ground truth and cleaned predictions, followed by greedy one-to-one matching within a 10-frame ($\approx$200 ms at 50 Hz) onset/offset tolerance, requiring at least 30\% overlap with the true event while prioritizing label agreement and tighter boundary alignment. To ensure consistency in metric calculation, the last 50 frames (1 s at 50 Hz) were masked to account for recordings ending with trailing silence.

Table~\ref{tab:pause detection results} summarizes per-type accuracies\remove{for each type}\remove{semantic (\textbf{\texttt{S}}) breathing (\textbf{\texttt{B}}), and breathing–semantic (\textbf{\texttt{BS}})pauses} together with overall accuracy across the three setups. The highest accuracies in each setup are highlighted in red. In setup \textcircled{1}, 1D CNN–LSTM with Emb-6 achieved the best overall accuracy (0.73) and highest on \textbf{\texttt{BS}} (0.86), while GRU with Emb-4 gave the strongest \textbf{\texttt{B}} (0.55) and VGG16 with Emb-4 reached strong \textbf{\texttt{S}} (0.80) but weak \textbf{\texttt{B}} (0.07).  \enlargethispage{0.5\baselineskip} In setup \textcircled{2}, fusion does not consistently improve performance: 1D CNN–LSTM with MFB+Emb-6 matched its single-feature MFB baseline (0.71) but fell short of Emb-6 alone, while MFCC+Emb-4 provided a balanced profile across \textbf{\texttt{S}} and \textbf{\texttt{BS}} (0.80/0.75). In setup \textcircled{3}, 1D CNN–LSTM with MFB+Emb-4 achieved the best overall accuracy in this block (0.72), 1D CNN–LSTM with MFCC+Emb-6 improved \textbf{\texttt{B}} to 0.54, and GRU with MFB+Emb-4 pushed \textbf{\texttt{S}} to 0.89 but with lower overall performance. These results show that (\emph{i}) mid-layer W2V2 embeddings (Emb-4/Emb-6) with 1D CNN–LSTM are strong single-feature baselines, (\emph{ii}) feature fusion is not uniformly beneficial, especially for \new{detecting} \textbf{\texttt{B}}\remove{ detection}, and (\emph{iii}) the two-stage pipeline improves \remove{semantic and breathing }detection \new{for \textbf{\texttt{S}} and \textbf{\texttt{B}}} while keeping overall accuracy competitive.

By pause type, \remove{semantic pauses}\new{detection for \textbf{\texttt{S}}} benefited from both acoustic features and embeddings: GRU with MFB and VGG16 with Emb-4 each achieved 0.80 accuracy, while the two-stage GRU with MFB+Emb-4 reached the highest at 0.89. \remove{Breathing pauses}\new{\textbf{\texttt{B}}} remained the most challenging: GRU with Emb-4 performed the best in the single-feature setup (0.55), and the two-stage 1D CNN–LSTM with MFCC+Emb-6 achieved 0.54, highlighting the value of acoustic–embedding fusion. \remove{Breathing–semantic pauses}\new{\textbf{\texttt{BS}}} were best captured by 1D CNN–LSTM with Emb-6 (0.86) and GRU with MFCC+Emb-4 and two-stages (0.85), suggesting embeddings consistently provide strong cues for mixed pauses.

Comparing embeddings, Emb-6 generally helped \remove{breathing detection}\new{detect \textbf{\texttt{B}}}, while Emb-4 often benefited \remove{semantic and mixed pauses}\new{\textbf{\texttt{S}} and \textbf{\texttt{BS}}}, especially with MFCC fusion. For overall accuracy, Emb-6 was favored with 1D CNN–LSTM + MFB fusion, while Emb-4 was stronger with two-stage (MFB) and GRU + MFCC fusion. Thus, Emb-6 was preferable when \textbf{\texttt{S}} or \textbf{\texttt{B}} was prioritized, while Emb-4 was better suited for \textbf{\texttt{BS}} detection, with model–feature pairing key to the best setup.

\new{For the two superior models, GRU and 1D CNN–LSTM, task–model fit matters:} GRU is more effective for \underline{\textit{classification}}, whereas 1D CNN–LSTM favors \underline{\textit{regression}}. A plausible explanation is inductive bias and capacity: the GRU’s lighter recurrent structure aligns well with discrete supervision but struggles to capture fine-grained continuous targets. By contrast, the 1D CNN–LSTM, which combines local spectral modeling with long-range context, is better suited to continuous regression of nuanced pause types. This suggests that regression provides a more compatible signal for higher-capacity models, while simpler GRUs benefit from categorical supervision. \new{For the two underperforming models, VGG16, with its large parameter count and stacked conv blocks, is calibrated for abundant, fine-grained data absent in our setting; AlexNet’s pooling-heavy, translation-invariant design blurs short temporal events and boundaries in MFB/MFCC and provides no explicit sequence modeling, yielding weak frame-wise separation of breathing, semantic, and co-occurring pauses.}


\noindent\textbf{Exertion-Level Classification:}
The best-performing baseline model and configuration in the case study of ~\cite{nie2025multi} was a 1D CNN-LSTM with W2V2 Emb-4, trained on spontaneous speech recordings, achieving an overall accuracy of $0.8102 \pm 0.04$. Under the current benchmark pipeline, the same combination was trained on only spontaneous speech and achieved accuracy of 0.7619 as marked in blue in Table~\ref{tab:benchmark}, and as well as on a combination of spontaneous speech and reading. As shown in Table~\ref{tab:benchmark}, spontaneous speech data alone achieved the best prediction accuracy of 0.9048, matching the same conclusion from ~\cite{nie2025multi}, and even outperformed the previous case study by 9.46$\%$. Model-wise, VGG16 performed poorly with W2V2 embeddings due to its fine-grained data-hungry multi-convolutional layer structure. AlexNet and GRU achieved the best accuracy of 0.9048 over the entire benchmark.  1D CNN-LSTM achieved the most stable performance across configurations with its two-type combined structure, with an average accuracy of 0.7922.

\section{Discussion and Future Work}
\new{A number of limitations can be identified for the proposed pipeline.} The filtered corpus used in this study, which is restricted to language fluency and treadmill exercises, narrows the demographic and linguistic diversity. The annotations for pause types also introduce human error, as distinctions between categories remain ambiguous despite concurrent data from the respiratory belt, aggravated by varying background noise and subject-microphone distance. The choice of fixed 15 s segmentation and $50\thinspace Hz$ label sampling rate simplifies training but loses resolution and introduces edge effects such as increased misclassifications near the beginning and end of the audio snippets. \new{Finally, our pre-set 4-class labels are limited, folding filled pauses, laughter, and coughs into \textit{other}.} \remove{Finally, our taxonomy focuses on \new{pre-set labels}\remove{{breathing, semantic, breathing+semantic, other}}, collapsing richer conversational phenomena such as filled pauses, laughter, or coughs into the residual \textit{other} class.} 

Future work for \textbf{Breathing and Semantic Pause Detection} could involve tailoring model–feature designs to specific pause types rather than using a uniform approach, \new{exploring alternative window lengths (e.g., 5s and 30s)}, applying more sophisticated fusion strategies,\remove{for example, combining multiple layers of W2V2 embeddings, may capture richer representations. Performance could be further improved by} leveraging multimodal inputs (e.g., PCG signals, subject demographic and fitness attributes), \remove{Training could also be augmented}\new{augmenting} with existing datasets (e.g., Sep-28k~\cite{lea2021sep}), and incorporating insights from linguistic analyses of human speech patterns which may help refine pause categorization. \new{Finally, we will explore quantization and pruning techniques to further optimize the framework, enhancing its real-time performance and deployability on edge devices.}

Future work for \textbf{Exertion-Level Classification} could extend to aligning exertion labels with acute behaviors induced by anaerobic exercise, and to deploying the pipeline on wearable or smart devices capable of collecting multimodal training data in gym or indoor settings~\cite{nie2022ai}. \remove{Additionally}\new{Further}, incorporating objective measurements of exertion as baseline references would enable systematic comparison with self-reported states, providing insight into when and how participants tend to over- or under-estimate their subjective perceptions in physiological studies.

\section{Conclusion}
This work provides a systematic exploratory study and evaluation of breathing and semantic pause detection and exertion-level classification in post-exercise speech, a setting that poses unique challenges due to irregular breathing patterns, overlapping events, and noise. Three major contributions include: (\emph{i}) a new annotation for the existing multimodal dataset, which fosters future research in health monitoring, sports coaching, and HCI. (\emph{ii}) An evaluation of three DL model setups on pause type detection with promising per-type accuracy up to 89\% for semantic, 55\% for breathing, 86\% for combined pauses, and 73\% overall. (\emph{iii})An exertion level classification setup provides an edge-ready tool to distinguish between cardio and anaerobic exercises with 90.48$\%$ accuracy. Our study underscores the potential of speech-based sensing as a non-invasive tool for monitoring physiological states, with implications for health monitoring, sports coaching, and human–computer interaction.

\vspace{-0.05in}
\bibliographystyle{ACM-Reference-Format}
\bibliography{references.bib}

\end{document}